# Analysis of Generative AI Policies in Computing Course Syllabi

Areej Ali
George Mason University
Fairfax, VA, USA
aali40@gmu.edu

Aayushi Hingle Collier
Montgomery College
Rockville, MD, USA
aayushi.hingle@montgomerycollege.edu

Umama Dewan
George Mason University
Fairfax, VA, USA
udewan@gmu.edu

Nora McDonald
George Mason University
Fairfax, VA, USA
nmcdona4@gmu.edu

Aditya Johri
George Mason University
Fairfax, VA, USA
johri@gmu.edu

## Abstract

Since the release of ChatGPT in 2022, Generative AI (GenAI) is increasingly being used in higher education computing classrooms across the United States. While scholars have looked at overall institutional guidance for the use of GenAI and reports have documented the response from schools in the form of broad guidance to instructors, we do not know what policies and practices instructors are *actually* adopting and how they are being communicated to students through course syllabi. To study instructors' policy guidance, we collected 98 computing course syllabi from 54 R1 institutions in the U.S. and studied the GenAI policies they adopted and the surrounding discourse. Our analysis shows that 1) most instructions related to GenAI use were as part of the academic integrity policy for the course and 2) most syllabi prohibited or restricted GenAI use, often warning students about the broader implications of using GenAI, e.g. lack of veracity, privacy risks, and hindering learning. Beyond this, there was wide variation in how instructors approached GenAI including a focus on how to cite GenAI use, conceptualizing GenAI as an assistant, often in an anthropomorphic manner, and mentioning specific GenAI tools for use. We discuss the implications of our findings and conclude with current best practices for instructors.

## 1 Introduction and Background

Since the introduction of ChatGPT in the Fall of 2022, most higher education institutions have responded with formal policies, often that encourage the use of GenAI and related applications [22]. Following the initial policies, that were often quite broad and simplistic, institutions have provided detailed guidance (including sample syllabi language) about range of use of GenAI [22], including within computing [21]. Yet, we do not know if, and how, these policies and guidance are actually being implemented by instructors. Furthermore, prior work found that schools are reticent about using GenAI in STEM related activities [22], so it is even less clear how these policies and guidance are being taken up by computing educators, an important concern given the potential of GenAI for impacting computing pedagogy [10].

While there is no clear agreement on how to best integrate GenAI into computing courses, several efforts are being made in that direction [27], and have some idea of computing educations' expectations for GenAI—both positive and negative—from research with pedagogical GenAI tools. For instance, research has explored ways that GenAI can improve computing learning outcomes [18], restore problem-based learning approaches, and transform the way instructors evaluate students [26]—the latter having a certain urgency given that GenAI has already been shown to produce passing marks in computer science course [31]. There is also some indication that the shift from writing to "reading, comprehending and evaluating" code will make teaching prompt engineering a vital skill but this will require thoughtful and deliberate integration into computing classrooms [9]. Computing educators may have reason for optimism, though, that some of the early failures of GenAI [23] might create learning opportunities that enlist students computational thinking [8]. For instance, research has explored how to use explanations to support computing learning [20].

At the same time, concerns about academic honesty, equity, and over-reliance on GenAI tools, and privacy can not easily be ignored both, in use of GenAI tools and in the code they produce [28, 32]. And keeping pace with changing GenAI technology [36] is decidedly daunting. As Prather et al. et al.'s meta review of existing literature in computing education and GenAI makes clear, GenAI presents opportunities to reduce instructor workload, on the one hand [26]. But there is effort associated with staying on top of changing technology, preparing new materials and adopting learning practices that are adaptive to those changes, on the other. Still, scholars argue that instructors must still engage with GenAI if they and their students are to benefit [2]. All of these tradeoffs might contribute to instructors willingness to (not) adopt GenAI in their courses and how they talk about it in their syllabi. In interviews with introductory computing professors, Lau and Guo found that most anticipated they would discourage its use, with some expressing a willingness to integrate it into their curriculum over the longer term, citing concerns about career opportunities [19]. To better understand instructors' policies and discourse about GenAI within their courses, we looked at computing syllabi from 54 R1 institutions in the U.S. and asked the following questions:

- **RQ1:** What GenAI policies are being adopted by instructors in their computing course syllabus?
- **RQ2.** What is the discourse around GenAI policies, specifically with regard to GenAI literacy and proficiency of use?

## 2 Methods
### 2.1 Assembling the syllabi corpus

Syllabi analysis is often used in computing education [13]. To create a corpus of institutions whose syllabi we could analyze, we used the Carnegie classification as the basis for data collection (an approach adopted by other scholars as well [4]). The Carnegie framework classifies doctoral-granting universities with high research activity as R1, of which there are 146 [1]. We focused our analysis on course syllabi that fall into the following categories: Software Development and Programming, Data Structures and Algorithms.

The researchers conducted a Google search for each university from our list using the following keyword search phrases were used: "[University name] computer science course syllabus ChatGPT spring 2024"; "[University name] computer science course syllabus generative AI spring 2024"; "[University name] computer science course syllabus artificial intelligence spring 2024"; "[University name] computer science course syllabus ChatGPT fall 2023"; "[University name] computer science course syllabus generative AI fall 2023"; "[University name] computer science course syllabus artificial intelligence fall 2023"; "[University name] computer science course syllabus ChatGPT"; "[University name] computer science course syllabus generative AI"; and "[University name] computer science course syllabus artificial intelligence." These search phrases resulted in 98 syllabi from 54 R1 Universities. Several syllabi (a total of N=6) for the same courses were collected and counted as distinct documents including: N=2 for 1 course with the same instructor from different time periods; N=2 syllabi for 1 course with different instructors and at the same time period; and N=2 syllabi for 1 course with different instructors and from different time periods.

We obtained the PDF copy and direct URL of each syllabus from *public sources* such as university repositories, course pages, and course instructors' GitHub pages. We excluded syllabi available through institutional resources such as private SharePoint sites or those posted on third-party websites (e.g., Chegg, Course Hero, Studocu, etc.) as those syllabi are not verified and privacy policies of those sites often preclude data collection. Data was collected from March 4, 2024 to May 11, 2024. Each syllabus was considered a distinct data unit and coded separately.

### 2.2 Analyzing the syllabi corpus

One researcher initially reviewed and open-coded the syllabi corpus and created a codebook. Two additional researchers reviewed and coded all 98 syllabi individually using the codebook. The three researchers then met to compare codes and reach agreement working through all 98 syllabi once again. See Figure 1 for the codes and relevant details. The final codebook consisted of 16 codes which are illustrated in Figure 1. These codes were then grouped into two themes that correspond to the two research questions. When reporting all codes, percentages are out of N=98. A syllabus can capture multiple subcodes for any given code and thus are reported out of 98 and not the total N for those codes. When we occasionally quote a section from a course syllabus, we do not provide formal references or links, as these are not published documents and instructors have reason to assume that they will not appear in publications. We do, however, provide links directly to syllabus deemed to represent some specified category of best practice (i.e., "exemplar syllabi" in the discussion).

## 3 Findings
### 3.1 RQ1: What policies are being adopted?

To answer RQ1, "What GenAI policies are being adopted by instructors in their computing course syllabus?", we grouped our syllabi based on four themes: *Range of use guidelines* describes the extent to which instructors permit the use of GenAI in class and assignments; *Explicit Encouragement/Discouragement* describes the way instructors communicated overtly their support for or disapproval of GenAI use; *Transparency* describes policies about GenAI citation and detection; and *Tools* describes the kinds of GenAI tools mentioned in syllabi (e.g., GenAI, CoPilot).

*3.1.1 Range of use guidelines.* Almost all (92%, N=90) syllabi provided explicit guidelines about permissions to use GenAI in their course, with half (50%, N=49) outright prohibiting use and a few only allowing it with explicit permission (N=6). About 41% (N=40) partially permitted use for activities specified in the syllabus. When explicit permission was required, instructors often stated that should cite GenAI use. Syllabi that *partially permitted* use tended to provide detailed breakdowns of the ways in which GenAI use was acceptable and not acceptable by students in the class.

*3.1.2 Explicit encouragement or discouragement.* Even while almost all syllabi had explicit policies about the use of GenAI, a few went further to communicate explicit encouragement and/or discouragement (17%, N=17). Syllabi were coded as explicitly *Discourages Use of GenAI Tools* (11%, N=11) and/or *Encourages Use of GenAI Tools* (7%, N=7) if the syllabus explicitly included the words "encourage" or "discourage", or if the language of the syllabus otherwise made it clear through expressions of overt enthusiasm (try it!) and/or references to approved or suggested use, or conversely, through reference to consequences for learning or for violating plagiarism policies, with emphasis (don't even think about it ...). These instances, whether positive or negative, tended to be more personal, where instructors took the time to make the point that they really encourage (often after anecdotally "using it") or discourage use of GenAI because of perceived *implications* (discussed in section 3.2).

*3.1.3 Transparency.* 83% (N=81) syllabi were coded as *transparency* because they provided explicit guidance on how to cite with citation guidelines, when to engage with AI, and/or referenced detection tools.

A little over two-thirds (N=65) stated that using GenAI without citation was in *violation* of the honor code and, in some cases, a violation of academic integrity on par with plagiarism. A little over a quarter of syllabi (N=27) mention the requirement that students use informal or formal (e.g., APA) *citation* of GenAI tools.

Often, when a syllabus had a clear academic integrity or honesty section, instructors included their GenAI policy and consequences for violating it.

Although not that common (12%, N=12), some did take a more involved approach, stipulating *additional requirements* that, if students used GenAI, they might have to provide detailed notation about how it was used and how a problem might have been solved

| Codes | Subcodes | N= | Definition |
|---|---|---|---|
| Range of Use Guidelines | GenAI Use Prohibited | 49 | Use of any GenAI tools is not allowed, this applies to any and all coursework. |
| | GenAI Use Permitted Partially | 40 | May be used for certain coursework and may not be allowed for other type of coursework. |
| | Requires Permission Before Use of GenAI Tools | 6 | If students intend to use GenAI tools, they must request and/or receive permission from an instructor before use in the course. |
| Explicit Encouragement or Discouragement | Encourages Use of GenAI Tools | 7 | Limited encouragement (i.e., encourages use as GenAI Assistant but not for other assignments). While some instructors may explicitly use the word "encourage," this may include the tone in a syllabus with the use of GenAI in the classroom. |
| | Discourages Use of GenAI Tools | 11 | Discourages to promote student learning experience or because the use of GenAI could violate academic integrity policies. While some instructors may explicitly use the word "discourage," this may include the tone in a syllabus with the use of GenAI in the classroom. |
| Transparency | GenAI Tools Citation/Disclosure | 27 | The syllabi specify the need for acknowledgment, annotations, informal citation, or formal citation (i.e., APA, MLA, Chicago, etc.) with the use of GenAI. |
| | Use of GenAI Detection Tools | 4 | If students decide to use GenAI, then instructors state that they may use GenAI detection tools. The policies explicitly state GenAI detection tools or refer to them, not just plagiarism detection in general. |
| | References Violation of Academic Integrity Policy | 65 | The use of GenAI could be seen as a violation of an honor code, academic integrity policy, etc. In some cases, a GenAI policy may be part of an academic integrity policy. |
| | Additional Requirements with Use of GenAI | 12 | Students utilizing GenAI tools may be subject to supplementary requirements or tasks to complete the course (i.e., provide additional explanations to show understanding of course concepts, provide detailed annotations, explain how a problem would have been solved without the use of GenAI, etc.). |
| Tools | Mention of Specific GenAI Tools | 91 | List or name specific GenAI tools such as ChatGPT, Bard, Llama, DALL-E, etc. |
| Implications | Skills Development with GenAI | 7 | Use of GenAI to develop prompt engineering skills for CS-related use cases. |
| | Hinders Learning | 17 | Use of GenAI tools may be detrimental to learning experience, building of foundational skills, possibly lead to difficulties in future careers, etc. |
| | Mentions Privacy and Awareness | 19 | Legal, privacy, security, and ethical implications. |
| | Mentions Veracity of GenAI Output | 20 | Instructors question whether GenAI output is accurate, biased, or misleading. The syllabus may also mention fact-checking GenAI output. |
| Anthropomorphism of GenAI | GenAI Assistant | 35 | Used to understand course concepts, study aid, etc. |
| | Characterization/ Personification of GenAI | 5 | Instructors appear to personify GenAI tools by directly comparing them to external entities that have abilities beyond general tools. In particular, instructors may prioritize the collaborative nature of GenAI technologies, presenting them as partners or team members who may have the ability to assist students beyond a typical tool. |

**Figure 1: Codes and Definitions**

without GenAI—essentially suggesting that there might, therefore, be no point in using it.

Very few (N=4) warned students that the school provide *detection tools*, or that they were at the instructor's disposal and discretion to use. When they did, however, they were somewhat emphatic. For example, one instructor warned students "Don't even think about turning in such work as your own, or even using it as a basis for your work. We have very sophisticated tools to find such cheating and we use them routinely. It's far better to get a 0 on an assignment (or exam) than to cheat."

*3.1.4 Tools.* Nearly all syllabi (N=91) mentioned specific GenAI tools and "ChatGPT" (N=90) was mentioned the most, followed by "CoPilot" (N=25). A few mentioned "Bard" (N=12), "Bing AI" (N=7), "GitHub CoPilot" (N=4), and "Grammarly" (N=2).

## 3.2 RQ2: What is the discourse Around GenAI Policies?

RQ2 asked, "What is the discourse around these GenAI policies, specifically regarding GenAI literacy and proficiency of use?" To answer this question, we identified two themes that comprise discourse: *Implications* which describes codes around perceived impact of GenAI on learning, skills, personal privacy, and society overall, and *Anthropomorphism of GenAI* which describes codes capturing the tendency to personify GenAI as an interactive assistant.

*3.2.1 Implications.* 34% (N=33) of syllabi mentioned implications for use of GenAI including the *veracity* of GenAI output (20%, N=20), concerns about *privacy* (19%, N=19), *hindering learning* (17%, N=17), and the ability to *develop skills* (7%, N=7).

When it came to concerns about *privacy*, some instructors emphasized the potential privacy violations associated with sharing personal information with OpenAI, as well as the legal or copyright violations associated with its outputs, including code. For example, one syllabus warned students that "code from AI tools that you do not understand could expose the university to loss of reputation or even financial harm through lawsuits." Instructors also warned students about GenAI *veracity* arguing that outputs from the tool are biased and misleading and sometimes inaccurate. Some syllabi suggested that students fact-check results. *Privacy* and *veracity* tended to coincide.

Almost as common (17%, N=17) was concern about *hindering learning* centering on the potential harm from use of GenAI to learning, sometimes suggesting that this would have implications for students' careers.

Fewer instructors talked explicitly about *skills development* with GenAI (7%, N=7). These instructors considered GenAI's benefits to developing prompt engineering skills for computer science-related use cases, and also as a way for students to develop skills with course materials. In other words, instructors seemed to considered using it the class as a way for students to both, get better at using GenAI and also to improve their engagement with class material and concepts.

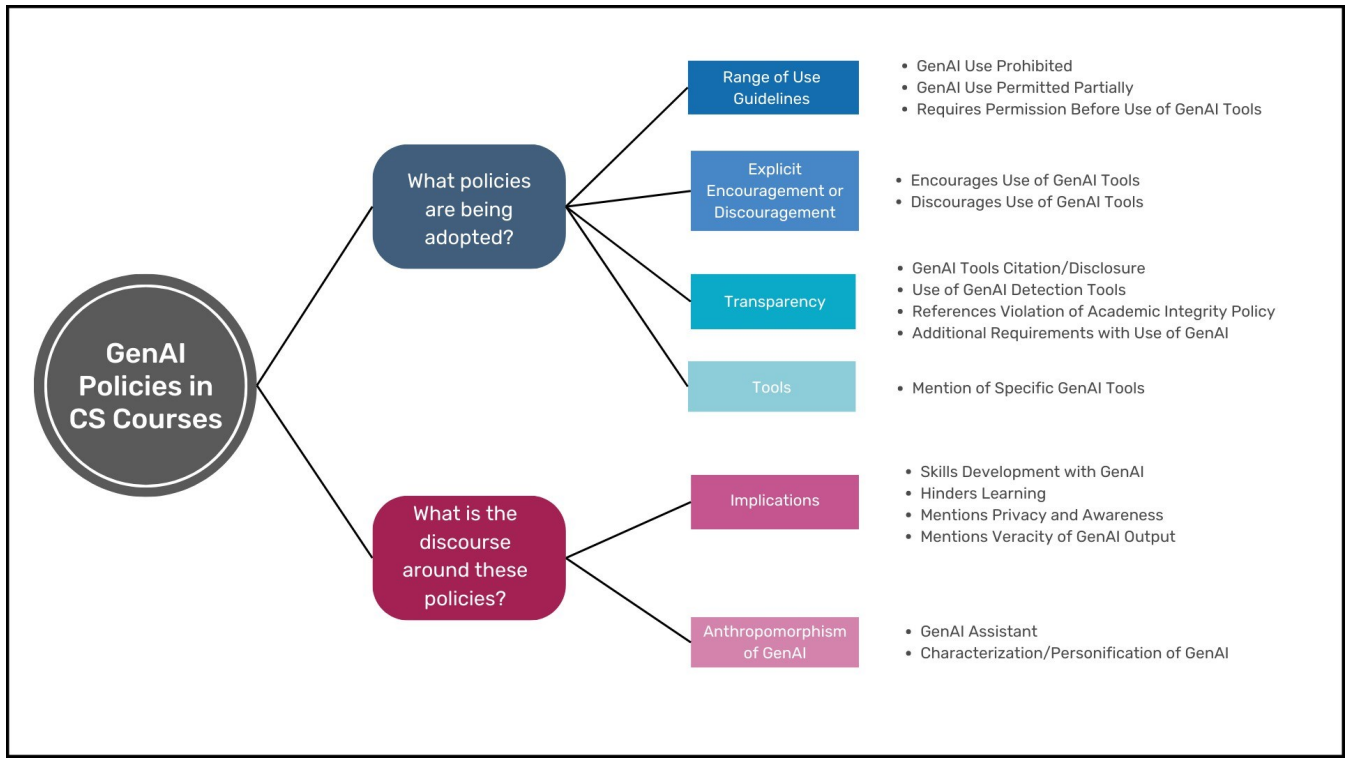

**Figure 2: Categories of Codes and Subcodes**

*3.2.2 Anthropomorphism of GenAI:.* 39% (N=38) syllabi were coded as *Anthropomorphism of GenAI* meaning they attributed living characteristics to GenAI, either as a *GenAI Assistant* (36%, N=35) or otherwise providing *Characterization/ Personification of GenAI* (5%, N=5). Below we describe these codes in more depth.

The *GenAI Assistant* code (36%, N=35) described GenAI as an "assistant" that helped students learn concepts by being, for example, a "study aid" that encouraged students to quiz them or deepen their understanding of concepts. Some even endorsed its veracity, for example: "I actually asked ChatGPT the question above and it came back with a pretty good answer! I suggest trying it!"

Although the *Characterization/Personification of GenAI* was less often captured (5%, N=5) it is nevertheless interesting because it describes a phenomena whereby instructors compared GenAI tools to people, such as "peers" or "tutors." For example, a data structures and algorithms course syllabus included a table that compared GenAI to human entities and specified how students can interact with it as part of the "collaboration policy." These instructors anthropomorphized GenAI as capable of generating seemingly advanced solutions to intricate problems or processes. They often implied that GenAI was a disruptive force that could alter traditional habits of thinking and problem-solving [16].

Several of these instructors prioritized the collaborative nature of GenAI technologies, presenting them as partners or team members who may assist students in improving their learning and creativity. These instructors insisted that students cite GenAI tools appropriately, sometimes going into much detail about the nuances of citing in these types of collaborations, the same way as they would someone else's work. But in thinking about these benefits, some imputed much ability to GenAI and worried aloud about GenAI's ability to exacerbate cheating.

## 3.3 Connections between policy and discourse:

*3.3.1 Encouragement and implications.* Syllabi that outright *encouraged use of GenAI tools* sometimes did not mention concern for its implications, like incorrect answers (e.g., *veracity*), *privacy violations*, and *hindering learning*. Indeed, instructors who encouraged use were more likely to mention the benefit of *skills development*. Those that promoted its use because of *skills development* also tended to promote the notion of AI as an *assistant*.

*3.3.2 Discouragement and implications.* By contrast, those that outright *discouraged* use of GenAI were more likely to provide *implications* like appealing to their students' knowledge of the *privacy* violation associated with using GenAI, the *veracity* or accuracy of the output of GenAI, and also how this can lead to a *diminished learning* for a real world situation, like a job interview, or a project where GenAI may not be accurate or useful.

*3.3.3 Transparency and implications.* Many syllabi in which instructors asked students to disclose use of, or cite GenAI also had statements communicating *privacy* or *veracity* concern with GenAI.

*3.3.4 Humans' that need to be cited and can (maybe) be trusted.* Instructors who communicated about GenAI as *anthropomorphized*

viewed engagement with it as a type of collaboration (with another entity) and the discourse reflected this. For instance, some instructors requested student's readily differentiate between what was their individual work and what was a product of collaboration, sometimes with excessive detail. Interestingly, those coded with *anthropomorphism* also tended to talk about the implications associated with GenAI *veracity* (or lack thereof). That is, for those who *anthropomorphize* GenAI, fact-checking was par for the course.

*3.3.5 Format for communicating policy.* Most syllabi included the policy on GenAI in the academic integrity part of the syllabus. But sometimes, a syllabus had a separate section where the instructor discussed GenAI use in their computing class, thereby merging policy with discourse in a more clear way.

## 4 Discussion

It's interesting that, even with a corpus of computing syllabi, we see the syllabi mostly only mention ChatGPT. For now, this seems to be the GenAI tool that instructors are most concerned with, and that is perhaps because they are still unfamiliar with what is available. Some have pointed out that lack of engagement with GenAI developments may limit their ability to adapt their courses appropriately [36]. There is indeed a danger of having ChatGPT be the stand-in for all GenAI, as different platforms have different capabilities or strengths, use different data, and have different politics.

Another interesting finding was that, in contrast to research of the formal policy statements and guidance by R1 universities—the majority of which encouraged use of GenAI—we see instructors tending slightly more often to prohibit or heavily restrict use [22]. This, in fact, consistent with research that suggests instructors (at least initially) expected to discourage use [19].

This is a challenging and rapidly evolving terrain. Our aim, in addition to reporting trends among computing educators, is to offer some guidance on best practices that we have seen in our survey of their computing syllabi.

### 4.1 Best practices: GenAI guidelines in syllabi

In our review, we identified syllabi that offered intriguing approaches, or what we deem "best practices." We highlight them below. Future research should investigate how instructors perceive of their success with these strategies.

**Parameters for explicit use.** A compelling syllabus format were those that were clear and detailed about when and how GenAI could be used. These syllabi were very explicit in the context of assignments, exams, etc. about when GenAI was (not) permitted and in what ways. A few detailed the parameters of permissions in a useful tabular format making it easy for students to understand while also communicating the importance of delineating use.

Sometimes a syllabus was specific about collaboration, but in way that still left us guessing. For instance, a few syllabi suggested using GenAI to solve coding problems but asked that students refrain from directly copying and pasting the output. While it is helpful to know that problem solving with GenAI is permitted, it's not entirely clear what the limits of this collaboration are for this course. In this case, some examples of the types of collaborative practices that are acceptable vs. not, might be helpful.

**Encouraging reflection.** It is, perhaps, also possible to help students think through the honesty and integrity of their actions. Syllabi that encouraged students to be self-reflecting and reflexive—thinking responsibly about use and assessing or fact-checking the output—seemed the most realistic in this regard. Placing that responsibility on students in an explicit way might lead students to practices that are consistent with what Moradi Dakhel et al. suggest where students learn from GenAI failures [23].

**Permissions and open lines of communication.** A few syllabi created openings for students to ask questions about use of GenAI. Syllabi that were coded as *requiring permission* before use, and that also had *additional requirements* around use, represent an important best practice. First, they keep channels of communication with students about their use of GenAI open. They normalize the behavior of checking in, allowing for students to get clarity about policies and for instructors to learn and adapt policies. Second, these policies essentially allow students to use GenAI but also may require them to do a computing problem on their own to satisfy the course requirements. This deters GenAI use by making it more work for students (e.g., because they have to do the problem themselves anyway). At the same time, this approach could promise greater engagement with the material (e.g., the requirement of doing the assignment with and without ChatGPT might enrich understanding of the problem or concept) as Moradi Dakhel et al. suggest [23]. This policy could therefore be a win-win; students use GenAI and get more deeper understanding or they don't use it and don't run the risk of violating academic honesty and integrity.

**Leveraging implications.** Syllabi that were explicit about the *implications* of GenAI sent a powerful message, particularly when they integrated implications with activities. An example of this would be when a syllabus provided meta commentary in the context of assignments about what, in fact, large language models are and what that means for the outputs they produce. A particularly good warning had to do with the veracity of ChatGPT where the syllabus stated that the tools trains on limited or out-of-date datasets. Another syllabus detailed the way that tools like ChatGPT create human-like responses that make them seem more credible than they are. This might discourage its use for dishonest work, while helping to facilitate healthy interaction where students question the outputs and, in the process, learn.

### 4.2 Exemplar Syllabi

To assist readers who might want to incorporate GenAI policies in their syllabus in this section we present a few exemplar syllabi that can serve as a model. A couple of key themes we identified are: use of authentic narrative and discussion of *implications* to frame guidance, personification of GenAI, and having students explain and justify use of GenAI. We have selected these through our review keeping in mind the breadth of what they capture and or specific policy items or examples that we judge to be exemplary.

The University of Chicago's *CMSC 22000: Introduction to Software Development* [5] course syllabus provides clear and explicit guidance about how to use GenAI in the course as well as a summary of GenAI *implications*—such as those associated with *veracity* and *hindering learning*—presented thoughtfully, with ethical scenarios that, for example, draw parallels between GenAI and a "classmate."

This is somewhat like the syllabus we describe later that take the approach of **personifying GenAI**.

Another notable and useful example is Syracuse University's *CIS352: Programming Languages Theory and Practice* [33]. This syllabus effectively breaks down what types of GenAI use are acceptable for different types of assignments. Adding to its effectiveness, is the way that the syllabus presents the instructor's own **authentic experience** with ChatGPT in a way that helps to lend credence to stated policies. Penn's *CIT 5950: Computer Systems Programming* [25] give students **authentic advice** based on *implications*. But in this case, the syllabus recommends not using GenAI. One important and nuanced argument made here involves articulation about concerns regarding *veracity*, which is that, when first learning about a topic the learner lacks expertise and might find it hard to make the right judgement about the information they receive from GenAI. Another notable trick that this syllabus employs is to say that when students use GenAI, they **should be prepared to explain their use of it in person during office hours**.

Dartmouth's *CS 10: Problem Solving via Object Oriented Programming* [7] course has one of the most comprehensive honor policies, with explanations and instructions for citations and the use of different GenAI tools. The syllabus instructs students that they must cite the GenAI tools used to complete an assignments, as well as be able to **explain every line of code**. The syllabus also provides some other detailed guidance, like (similar to [25]) asking students to not share their prompts or outputs with others, and to not to use code related to a concept that has not yet been covered in class.

Princeton's *COS226: Data Structures and Algorithms* [29] syllabus usefully details what is acceptable use of GenAI, and what is not, through **personification of GenAI** as "AI chatbots" in a table where all other types of human "collaborators" are listed. The **formatting** of this syllabus was exemplar.

The University of Massachusetts Amherst's *CICS 110: Foundation of Programming* [34] course syllabus is notable for including an **"AI Assistants"** section that discusses how GenAI should be used as a learning tool but "not a substitute for [student's] own efforts." The syllabus emphasizes *implications* in terms of *hindering learning*.

Finally, we want to point interested readers in other exemplar resources such as those from centers of teaching and learning [14], and sites maintained by individuals [11].

### 4.3 Implications

**Implications for Teaching, Pedagogy and Assessment.** As our analysis shows, interest in GenAI within course syllabi is still largely limited to concerns with plagiarism and academic integrity violations [3]. Instructors for the most part have not incorporated the use of GenAI for instruction and learning in a designed manner in their courses. Based on the data, we found that there is an opportunity to use GenAI itself as a "teaching moment" and talk about how AI and related technologies work, their pros and cons, and their equitable implementation. There is also an opportunity here to increase students' "learning ecology" by providing guidance to use GenAI fruitfully in their learning process (e.g., as a tutor or coach) [35]. This, of course, will require significantly re-designing some courses and their implementations, and could change approaches to assessment, as others have also pointed out [24]. So much of the effort of students often goes into bypassing or gaming certain aspects of an assessment rather than learning and this long-term concern has persisted [15]. Finally, workforce development is a concern as well in the use of GenAI in computing courses given that GenAI use in the industry is on the rise. There is balance to be struck between banning students from using GenAI and encouraging and supporting them, through skills such as prompting [30], so long as learning goals are achieved [6, 12]. There is also a need to train students on GenAI literacy given that many students have misconceptions about how it works [17].

**Implications for Future Research.** As a next step it is important to undertake this analysis on a larger dataset more representative of other kinds of institutions and programs, and CS-allied disciplines. More representative data and analysis can help establish best practices that are more usable and generalizable. Furthermore, gaining a better understanding of the outcomes for different GenAI policies on student learning is another critical empirical goal. Another approach is to better understand how faculty see GenAI, i.e., what mental models do they have? Currently, a lot of assumptions are being made about faculty knowledge and expertise but it might be prudent to identify opportunities for faculty training.

## 5 Conclusion and Limitations

This paper reports on the policies and discourse surrounding the use of GenAI in higher education computing classrooms. We found that syllabi policy include a range of prohibitions and permissions. Instructors are reluctant to wholly embrace GenAI and seem to be not as enthusiastic as their universities in embracing GenAI. The strongest syllabi are those that are clear about the implications of its use and have nuanced policies about how and when to use it. Future work should extend the data corpus to other institutions, in particular teaching-centered universities and colleges, so that a more balanced perspective can be developed.

### Acknowledgments

This work is partly supported by US NSF Awards 2319137, 1954556, and USDA/NIFA Award 2021-67021-35329. Any opinions, findings, and conclusions or recommendations expressed in this material are those of the authors and do not necessarily reflect the views of the funding agencies.